\input amstex
\magnification=1200
\loadbold
\loadeufm
\loadmsbm
\vsize=18.true cm
\hsize=13.5true cm

\TagsOnRight

\def\Im{\operatorname{Im}}
\def\Re{\operatorname{Re}}

\def\Tr{\operatorname{Tr}}

\def\notni{\hbox{{\kern0pt\raise.7pt\hbox{${\scriptstyle +}$}}
{\kern-9.5pt\raise0pt\hbox{$\supset$}}}}

\font\sbf = cmbx10 scaled\magstep2

\centerline{\sbf IRREDUCIBLE RESPESENTATIONS}
\medskip
\centerline{\sbf OF CAYLEY-KLEIN UNITARY}
\medskip
\centerline{\sbf  ALGEBRAS}

\bigskip
\centerline{\bf N.A.~Gromov}

\medskip
\centerline{Mathematical Department of Komi Research Center}
\centerline{of Russian Academy of Sciences,}
\centerline{Kommunisticheskaja Str. 24, Syktyvkar,}
\centerline{Komi Republic, Russia}
\centerline{E-mail: gromov\@cbis.komi.su}

\medskip
\centerline{\bf S.S.~Moskaliuk}

\medskip
\centerline{Bogoljubov Institute for Theoretical Physics}
\centerline{of National Academy of Sciences of Ukraine,}
\centerline{Metrologichna Str. 14-b}
\centerline{252143, Kiev-143, Ukraine}
\centerline{E-mail: mss\@gluk.apc.org}

\vskip 2truecm

\centerline{\bf Abstract}\medskip

Multidimensional contractions of irreducible representations of
 the Cayley-Klein unitary algebras in the Gel'fand-Zetlin basis are
considered.   Contracted over different parameters, algebras
can turn
out to be isomorphic. In this case method of transitions describes the
same reducible representations in different basises, say, discrete and
continuous ones.

\vfill\eject

{\bf 1. Representation of unitary algebras $u\,(2;j_1)$}
\medskip
{\bf 1.1. Finite-dimensional irreducible representations }
{\bf of algebra $u\,(2)$}
\medskip

These representations have been described by Gel'fand and Zetlin [1]. They
are realized in the space with orthogonal basis, determined by a
scheme with integer-valued components
$$
|\,m^*\rangle=\left.\left| \matrix
m_{12}^*\ \ m^*_{22}\\\vspace{1\jot}
m^*_{11}
\endmatrix\right.
\right>,\quad m^*_{12}\geq m^*_{11}\geq m^*_{22},
\tag 1
$$
by operators
$$
\spreadlines{1\jot}
\gathered
E^*_{11}\,|\,m^*\rangle=m^*_{11}\,|\,m^*\rangle=
A^*_{11}\,|\,m^*\rangle,\\
E^*_{22}\,|\,m^*\rangle=(m^*_{12}+m^*_{22}-m^*_{11})\,|\,m^*\rangle=
A^*_{00}\,|\,m^*\rangle,\\
E^*_{21}\,|\,m^*\rangle=\sqrt{(m^*_{12}-m^*_{11}+1)(m^*_{11}-m^*_{22})}
\,|\,m^*_{11}-1\rangle=\\
=A^*_{01}\,|\,m^*\rangle,\\
E^*_{12}\,|\,m^*\rangle=\sqrt{(m^*_{12}-m^*_{11})(m^*_{11}+1-m^*_{22})}\,|\,
m^*_{11}+1\rangle=\\
=A^*_{10}\,|\,m^*\rangle,
\endgathered
\tag 2
$$
where
$|\,m^*_{11}\pm1\rangle$
means the scheme (1) with the component $m^*_{11}$ changed for
$m^*_{11}\pm1$.

Let us change standard notations of generators $E_{kr}$ for new
notations
$A_{n-k,n-r}$, $n=2$, consistent with the notations of [2]. The
irreducible representation is completely determined by the components
$m^*_{12}$, $m^*_{22}$, $m^*_{12}\geq m^*_{22}$
of the upper row in (1) (components of the major weight).

As it is known, Casimir operators are proportional to the unit operators on
the space of irreducible representation. The \pagebreak spectrum of
Casimir operators for classical groups was found in [3-5]; for
semisimple groups and algebra $u\,(2)$ it is as follows:
$$ C^*_1=m^*_{12}+m^*_{22},\quad
C^*_2=m^{*2}_{12}+m^{*2}_{22}+m^*_{12} -m^*_{22}.
\tag 3 $$
Let us remind that the asterisk marks the quantities referring to the
classical groups (algebras).

In the space of representation there is vector of the major weight
$\varphi_{\scriptscriptstyle\text{Mw}}$, described by the scheme (1)
for $m^*_{11}=m^*_{12}$. Acting on it, the rising operator $A^*_{10}$
gives zero $(A^*_{10}\varphi_{\scriptscriptstyle\text{Mw}}=0)$ and the
lowering operator $A^*_{01}$ makes the value $m^*_{11}=m^*_{12}$ less
by one
$(A^*_{01}\varphi_{\scriptscriptstyle\text{Mw}}=
\sqrt{m^*_{12}-m^*_{22}}\,|\,m^*_{12}-1\rangle)$.
Consequently applying $A^*_{01}$ to
$\varphi_{\scriptscriptstyle\text{Mw}}$, we come to the vector of
the minor weight $\varphi_{\scriptscriptstyle\text{mw}}$, described by
the scheme (1) for $m^*_{11}=m^*_{22}$. Acting on
$\varphi_{\scriptscriptstyle\text{mw}}$, the lowering
operator gives zero
$(A^*_{01}\varphi_{\scriptscriptstyle\text{mw}}=0)$. The irreducible
representations is finite-dimensional, and this fact is reflected in
the inequalities (1), which are satisfied by the component
$m^*_{11}$ of the scheme.

The condition of unitary for representations of the algebra $u_2$ is
\linebreak equivalent to the following relations for the operators (3):
$A^*_{kk}=\bar{A}^*_{kk}$ $(k=0,1)$, $A^*_{01}=\bar{A}^*_{10}$,
where the bar means the complex conjugation. For matrix elements the
conjugation of unitary can be written as follows
$$
\gathered
\langle m^*\,|\,A^*_{00}\,|\,m^*\rangle=
\overline{\langle m^*\,|\,A^*_{00} \,|\,m^*\rangle},\\
\langle m^*\,|\,A^*_{11}\,|\,m^*\rangle=
\overline{\langle m^*\,|\,A^*_{11} \,|\,m^*\rangle},\\
\langle m^*_{11}-1\,|\,A^*_{01}\,|\,m^*\rangle=
\overline{\langle m^*\,|\,A^*_{10}\,|\,m^*_{11}-1\rangle}.
\endgathered
\tag 4
$$
\bigskip

{\bf 1.2. Transition to the representations of algebra $u\,(2;j_1)$}
\medskip

Under transition from the algebra $u\,(2)$ to the algebra $u\,(2;j_1)$
the generators $A^*_{00}$, $A^*_{11}$ and the Casimir operators
$C^*_1$ remain unchanged, and the generators $A^*_{01}$, $A^*_{10}$
and the Casimir operator $C^*_2$ are transformed as follows (see [2]):
$$
A_{01}=j_1A^*_{01}(\to),\quad
A_{10}=j_1A^*_{10}(\to),\quad
C_2(j_1)=j_1^2C^*_2(\to),
\tag 5
$$
where
$A^*_{01}(\to)$, $A^*_{10}(\to)$ are singularly transformed (for dual
value of parameter $j_1=\iota_1$) generators of the initial algebra
$u\,(2)$. A \pagebreak question is now raised as to how to set this
transformation for the irreducible representation (2) of the algebra
$u\,(2)$. Let us give the transformation of the components of scheme
(1) as follows:
$$
m_{12}=j_1m^*_{12},\quad
m_{22}=j_1m^*_{22},\quad
m_{11}=m^*_{11}.
\tag 6
$$
Then, taking into account (5), the
representation generators (2) can be written as
$$
\gathered
A_{00}\,|\,m\rangle=\left.\left.\biggl({m_{12}+m_{22}\over
j_1}-m_{11}\biggr)\right|m\right>,\quad
A_{11}\,|\,m\rangle=m_{11}\,|\,m\rangle,\\
A_{01}\,|\,m\rangle=\sqrt{[m_{12}-j_1(m_{11}-1)](j_1m_{11}-m_{22})}
\,|\,m_{11}-1\rangle,\\
A_{10}\,|\,m\rangle=\sqrt{(m_{12}-j_1m_{11})[j_1(m_{11}+1)-m_{22}]}
\,|\,m_{11}+1\rangle,
\endgathered
\tag 7
$$
and the spectrum of Casimir operators
$$
\gathered
C_1(j_1)=A_{00}+A_{11},\\
C_2(j_1)=A_{01}A_{10}+A_{10}A_{01}+j_1^2(A^2_{00}+A^2_{11})
\endgathered
\tag 8
$$
are
$$
C_1(j_1)={m_{12}+m_{22}\over j_1},\quad
C_2(j_1)=m^2_{12}+m^2_{22}+j_1(m_{12}-m_{22}),
\tag 9
$$
where $|\,m\rangle$ means the following scheme
$$
|\,m\rangle=
\left.\left| \matrix
m_{12}\ \ m_{22}\\\vspace{1\jot}
m_{11}
\endmatrix\right.
\right>.
\tag 10
$$
The
inequality (1) for components can be formally written as
$$
{m_{12}\over j_1}\geq m_{11}\geq {m_{22}\over j_1},\quad
{m_{12}\over j_1}\geq {m_{22}\over j_1}.
\tag 11
$$

To reveal the sense of these inequalities for $j_1\neq1$, we shall
discuss the action of rising operator $A_{10}$ on the vector of the
``major weight'' $\varphi_{\scriptscriptstyle\text{Mw}}$, described
by scheme (10) for $m_{11}=m_{12}$, and the \pagebreak lowering
operator $A_{01}$ on the vector of the ``minor weight''
$\varphi_{\scriptscriptstyle\text{mw}}$, described by scheme
(10) for $m_{11}=m_{22}$. We obtain
$$
\gathered
A_{10}\varphi_{\scriptscriptstyle\text{Mw}}=\sqrt{m_{12}(1-j_1)[j_1
(m_{12}+1)-m_{22}]}\,|\,m_{12}+1\rangle,\\
A_{01}\varphi_{\scriptscriptstyle\text{mw}}=\sqrt{[m_{12}-j_1
(m_{22}-1)]m_{22}(j_1-1)}\,|\,m_{22}-1\rangle.
\endgathered
\tag 12
$$
It can be seen from here that for $j_1=\iota_1,i$
these expressions differ from zero.  Therefore, the space of
representation is infinite-dimensional, and the integer-valued
component $m_{11}$, which numbers the basis vectors, varying from
$-\infty$ to $\infty$. Thus, the formal inequalities (11) for
$j_1=\iota_1,i$ are interpreted as
$\infty>m_{11}>-\infty$ and $m_{12}\geq m_{22}$.

The form (6) of the transformation of Gel'fand-Zetlin scheme is
chosen in such a way that the Casimir operator of the second order
would differ from zero and not contain indeterminate expressions for
$j_1=\iota_1$.
\bigskip

{\bf 1.3. Contractions of irreducible representations}
\medskip

For $j_1=\iota_1$ the operator $A_{00}$ contains the summand
$(m_{12}+m_{22})/\iota_1$, which is in general, undeterminate, if its
numerator is a real, complex or dual number. This summand is
determinate, if its numerator is purely dual number (see [6])
$m_{12}+m_{22}=\iota_1\zeta$, where $\zeta\in\Bbb R$ or
$\zeta\in\Bbb C$. The requirement of unitarity for the operator
$A_{00}$ is given by $\zeta\in\Bbb R$. Thus, in order that the
operators (7) would determine the representation of the algebra
$u\,(2;\iota_1)$, it is necessary to choose the components $m_{12}$,
$m_{22}$ of the scheme (9) as follows
$$
m_{12}=k+\iota_1\zeta/2,\quad m_{22}=-k+\iota_1\zeta/2,\quad
\zeta\in\Bbb R,
\tag 13
$$
where $k$, generally
speaking, is a complex number.

The scheme (10)  for dual values of components is determined by the
expansion into series
$$
|\,m\rangle=\left.\left|\matrix
k+\iota_1\zeta/2-k+\iota_1\zeta/2\\
m_{11}\endmatrix
\right.\right>=|\,\widetilde{m}\rangle+\iota_1{\zeta\over2}(\,|\,
\widetilde{m}\rangle'_{12}+\,|\,\widetilde{m}\rangle'_{22}),
\tag 14
$$
where
$|\,\widetilde{m}\rangle=\left.\left|\matrix
k-k\\
m_{11}\endmatrix
\right.\right>$, $|\,\widetilde{m}\rangle'_{12}={\partial\over
\partial m_{12}}\,|\,m\rangle\,|\,_{m_{12}=k,\ m_{22}=-k}$,
and the similar expressions are valid for
$|\,\widetilde{m}\rangle'_{22}$.
The initial schemes (1) are normalized to unit:
$\langle m^{*'}\,|\,m^*\rangle=\delta_{m^{*'}_{21}m^*_{22}}
\delta_{m^{*'}_{22}m^*_{22}}\delta_{m^{*'}_{11}m^*_{11}}$.
Schemes (10) \pagebreak for the
continuous values of components are normalized to delta-function. In
particular, for $|\,\widetilde{m}\rangle$ we have normalization to the
squared delta-function
$$
\langle\widetilde{m}'\,|\,\widetilde{m}\rangle=\delta^2(k'-k)\delta_{m'_{11}
m_{11}}.
\tag 15
$$

Substituting (13), (14) in formulas of $\S$1.2, we obtain the
representation operators of the algebra $u\,(2;\iota_1)$ (the dual
parts are omitted):
$$
\gathered
A_{00}\,|\,\widetilde{m}\rangle=(\zeta-m_{11})\,|\,\widetilde{m}\rangle,
\quad A_{11}\,|\,\widetilde{m}\rangle=m_{11}\,|\,\widetilde{m}\rangle,\\
A_{01}\,|\,\widetilde{m}\rangle=k\,|\,\widetilde{m_{11}-1}\rangle,
\quad A_{10}\,|\,\widetilde{m}\rangle=k\,|\,\widetilde{m_{11}+1}\rangle.
\endgathered
\tag 16
$$
The requirement of unitarity (4) for
operators $A_{01}$, $A_{10}$ gives $k=\bar{k}$, i.e. $k$ is a real
number, the inequality $m_{12}\geq m_{22}$ gives for the real parts
$k\geq-k$, i.e. $k\geq0$, the component $m_{11}$ is integer-valued and
changes according to (10) in the range $-\infty<m_{11}<\infty$.  The
eigenvalues of Casimir operators (8) on the irreducible
representations
of the algebra $u\,(2;\iota_1)$ are
$$
C_1(\iota_1)=\zeta,\quad C_2(\iota_1)=2k^2.
\tag 17
$$
They are independent and differ from zero. As in the case of the
initial algebra $u\,(2)$, the irreducible representations of the
contracted algebra $u\,(2;\iota_1)$ are completely determined by the
upper row of the scheme, i.e. by parameters $k\geq0$, $\zeta\in\Bbb
R$.  The results (14), (15) coincide with the corresponding
formulas in [7] for the case of algebra $iu\,(1)$.

To the requirement of determinacy of the spectrum of operator
$C_2(\iota_1)$ corresponds not only the transformation (6) of the
components of Gel'\-fand-Zetlin schemes, but, for example, the
transformation
$m_{12}=j_1m^*_{12}$, $m_{22}=m^*_{22}$, $m_{11}=m^*_{11}$
as well. Here generator
$A_{00}\,|\,m\rangle=({m_{12}\over\iota_1}+m_{22}-
\mathbreak-m_{11})
\,|\,m\rangle$
is determined only for
$m_{12}=\iota_1p$, $p\in\Bbb R$, but then
$C_1(\iota_1)=\mathbreak=p+m_{22}\neq0$, and
$C_2(\iota_1)=\iota_1^2[m_{22}(m_{22}
-1)+m_{12}^2/\iota_1^2+m_{12}/\iota_1]=m_{12}^2+\mathbreak+\iota_1m_{12}
=(\iota_1p)^2+\iota_1(\iota_1p)=0$.  In this case the irreducible
representation of the algebra $u\,(2)$ is contracted to the degenerate
representation of the algebra $u\,(2;\iota_1)$, for which
$C_1(\iota_1)\neq0$, and $C_2(\iota_1)=0$. One can not, at all,
transform the components $m_{kr}=m^*_{kr}$. Then \pagebreak under
contraction we also obtain the degenerate representation of the
algebra $u\,(2;\iota_1)$ with $C_1(\iota_1)=m_{12}+m_{22}\neq0$,
$C_2(\iota_1)=0$.  This representation is given by
generators $A_{00}$, $A_{11}$ of the form (2), and the generators
$A_{01}$ and $A_{10}$ bring $|\,m\rangle$ to zero:
$A_{01}\,|\,m\rangle=0$, $A_{10}\,|\,m\rangle=0$.

We have chosen the transformation (6) which gives under contraction
the non-degenerated general representation of the algebra
$u\,(2;\iota_1)$ with non-zero spectrum of all Casimir operators.
Further, studying the algebras of the higher dimensions, we shall
consider just this case.
\bigskip

{\bf 1.4. Analytical continuation of irreducible representations}
\medskip

As it has been noticed in [6], the formulas for the
transformation of algebraic quantities, derived from the requirement
of the absence of undeterminate expressions for dual values of
parameters $\boldkey j$, are valid for imaginary values  of parameters
as well. For the algebra $u\,(2;j_1=i)\equiv u\,(1,1)$
this means that $(m_{12}+m_{22})/i=\zeta$. The requirement of
unitarity for $A_{00}$ gives $\zeta\in\Bbb R$, i.e. components
$m_{12}$ and $m_{22}$, in general, are $$
m_{12}=a+i\biggl(b+{\zeta\over2}\biggr),\quad
m_{22}=-a-i\biggl(b-{\zeta\over2}\biggr),\quad a,b,\zeta\in\Bbb R.
\tag 18
$$
Substituting (18) in (9), (7), we get
$$
\gather
A_{00}\,|\,m\rangle=(\zeta-m_{11})\,|\,m\rangle,\quad
A_{11}\,|\,m\rangle=m_{11}\,|\,m\rangle,\\ \vspace{1\jot}
A_{01}\,|\,m\rangle=\\ \vspace{1\jot}
=\sqrt{a^2-b(b+1)+\biggl({\zeta\over2}-m_{11}\biggr)
\biggl({\zeta\over2}-m_{11}+1\biggr)+}\dddot{\phantom{\biggl(
\biggr)^A}}\overrightarrow{\phantom{\biggl(\biggr)^A}}\\ \vspace{1\jot}
\overleftarrow{\phantom{b}}\dddot{\phantom{b}}
\overline{+ia(2b+1)}\,|\,m_{11}-1\rangle,\\ \vspace{1\jot}
A_{10}\,|\,m\rangle=
\tag 19 \\
\vspace{1\jot}
=\sqrt{a^2-b(b+1)+\biggl({\zeta\over2}-m_{11}\biggr)
\biggl({\zeta\over2}-m_{11}-1\biggr)+}
\dddot{\phantom{\biggl(
\biggr)^A}}\overrightarrow{\phantom{\biggl(\biggr)^A}}
\endgather
$$

$$
\gathered
\overleftarrow{\phantom{b}}\dddot{\phantom{b}}
\overline{+ia(2b+1)}\,|\,m_{11}+1\rangle,\\ \vspace{1\jot}
C_1(i)=\zeta,\quad C_2(i)=2\left[a^2-b(b+1)-
\biggl({\zeta\over2}\biggr)^2\right]+2ia(2b+1).
\endgathered
$$

The relation (4) for the operators $A_{01}$, $A_{10}$, implied by
the requirement of Hermiticity, can be written as follows
$$
\gathered
\sqrt{a^2-b(b+1)+\biggl({\zeta\over2}-m_{11}\biggr)\biggl({\zeta\over2}-
m_{11}+1\biggr)+ia(2b+1)}=\\
=\sqrt{a^2-b(b+1)+\biggl({\zeta\over2}-m_{11}\biggr)\biggl({\zeta\over2}-
m_{11}+1\biggr)-ia(2b+1)}.
\endgathered
\tag 20
$$

To satisfy (20) for any $\zeta$, $m_{11}$, the imaginary part of the
radicand must vanish and the real part must be positive. It is
possible in two cases: $a)$
$b=-{1\over2}$, $a\neq0$;
$b)$ $a=0$, $-b(b+1)>0$.

In the case $(a)$ the formulas (19) can be rewritten as follows:
$$
\gathered
A_{01}\,|\,m\rangle=\sqrt{a^2+[m_{11}-(1-\zeta)/2]^2}\,|\,m_{11}-1\rangle,\\
A_{10}\,|\,m\rangle=\sqrt{a^2+[m_{11}+(1-\zeta)/2]^2}\,|\,m_{11}+1\rangle,\\
C_1(i)=\zeta,\quad C_2(i)=2a^2+(1-\zeta^2)/2.
\endgathered
\tag 21
$$

This is irreducible representation of the continuous series of the
algebra $u\,(1,1)$.  Gel'fand and Graev [8] used the components
$\widetilde{m}_{12}=-{1\over2}+\sigma$,
$\widetilde{m}_{22}={1\over2}+\sigma$,
related with components $m_{12}=i\widetilde{m}_{12}$,
$m_{22}=i\widetilde{m}_{22}$ via formulas $\Re\sigma={\zeta\over2}$,
$\Im\sigma=-a$.

In the case $(b)$ the relations (19) can be rewritten as follows:
$$
\gathered
A_{01}\,|\,m\rangle=\sqrt{[m_{11}-(\zeta+1)/2]^2-\biggl(
b+{1\over2}\biggr)^2}\,|\,m_{11}-1\rangle,\\
A_{10}\,|\,m\rangle=\sqrt{[m_{11}-(\zeta-1)/2]^2-\biggl(
b+{1\over2}\biggr)^2}\,|\,m_{11}+1\rangle,\\
C_1(i)=\zeta,\quad C_2(i)=-2\left[b(b+1)+{1\over4}\zeta^2\right].
\endgathered
\tag 22
$$
This is irreducible representation of additional continuous series [9].

There is once more possibility besides cases $(a)$ and $(b)$. Let
components $m_{12}$, $m_{22}$ be purely imaginary:
$m_{12}=i\widetilde{m}_{12}$, $m_{22}=i\widetilde{m}_{22}$, where
$\widetilde{m}_{12}$, $\widetilde{m}_{22}$ are integers. Then the
relations (19) can be rewritten as follows:
$$
\gathered
A_{00}\,|\,m\rangle=(\widetilde{m}_{12}+\widetilde{m}_{22}-m_{11})
\,|\,m\rangle,\quad A_{11}\,|\,m\rangle=m_{11}\,|\,m\rangle,\\
A_{01}\,|\,m\rangle=\sqrt{-(\widetilde{m}_{12}-m_{11}+1)
(m_{11}-\widetilde{m}_{22})}\,|\,m_{11}-1\rangle,\\
A_{10}\,|\,m\rangle=\sqrt{-(\widetilde{m}_{12}-m_{11})
(m_{11}+1-\widetilde{m}_{22})}\,|\,m_{11}+1\rangle,\\
C_1(i)=\widetilde{m}_{12}+\widetilde{m}_{22},\quad
C_2(i)=-(\widetilde{m}_{12}^2+\widetilde{m}_{22}^2+\widetilde{m}_{12}
-\widetilde{m}_{22}),
\endgathered
\tag 23
$$
They coincide with (2), (3) except for the sign minus in the
radicand. The requirement of unitarity (4) can be reduced to the
reality of the root in the expressions for the generators $A_{01}$ and
$A_{10}$ which is possible when one of the factors is negative. As a
result, we get two more irreducible representations: $c)$
$m_{11}\geq\widetilde{m}_{12}+1$; $d)$
$m_{11}\leq\widetilde{m}_{22}-1$, which are called discrete series.
The discrete series of irreducible representations of pseudounitary
algebras $u\,(p,q)$ are described by Gel'fand and Graev [8, 10].
The cases $(c)$ and $(d)$ correspond to modified schemes
$$
\left|\left.\matrix
&\widetilde{m}_{12}&\widetilde{m}_{22}\\
m_{11}&&\endmatrix
\right.\right>,\quad
\left|\left.\matrix
\widetilde{m}_{12}&\widetilde{m}_{22}&\\
&&m_{11}\endmatrix
\right.\right>.
\tag 24
$$

In the simplest case of algebras $u\,(2;j_1)$ we have shown in detail
how method of transitions works for irreducible representations.

The irreducible representations of algebras $u\,(2;j_1)$ are given by
formulas of $\S$1.2 with additional conditions (13) in the case of
contraction and (18) in the case of analytical continuation to the
components of the upper row in Gel'fand-Zetlin scheme. To obtain
unitary representation, it is necessary additionally to check up
whether the relations (4) are satisfied for contracted and
analytically continued generators of representation.
\vfill\eject

{\bf 2. Representations of unitary algebras $u\,(3;j_1,j_2)$}
\medskip

{\bf 2.1. Description of representations}
\medskip

Standard notations of Gel'fand and Zetlin [11] correspond to
diminishing chain of subalgebras $u\,(3)\supset u\,(2)\supset u\,(1)$,
where $u\,(3)=\{E_{kr},\quad k,r=1,2,3\}$;
$u\,(2)=\{E_{kr},\quad k,r=1,2\}$; $u\,(1)=\{E_{11}\}$. To make them
consistent with the notations of [2], it is necessary to change
index $k$ for index $n-k=3-k$, i.e. $E_{kr}=A_{n-k,n-r}$. Doing so, we
turn the chain of subalgebras into $u\,(3;j_1,j_2)\supset
u\,(2;j_2)\supset u\,(1)$, where $u\,(3;\boldkey j)=\{A_{sp},\quad
s,p=0,1,2\}$; $u\,(2;j_2)=\{A_{sp},\quad s,p=1,2\}$;
$u\,(1)=\{A_{22}\}$. The component enumeration in Gel'fand-Zetlin
schemes we leave unchanged.

It is well known that to determine representations of algebra $u\,(3)$
it is sufficient to determine action of generators $E_{pp}$,
$E_{p,p-1}$, $E_{p-1,p}$, i.e. generators $A_{kk}$ $(k=0,1,2)$,
$A_{k,k+1}$, $A_{k+1,k}$ $(k=0,1)$. The rest generators $A_{02}$,
$A_{20}$ can be found using commutators $A_{02}=[A_{01},A_{12}]$,
$A_{20}=[A_{21},A_{10}]$. Under transition from $u\,(3)$ to
$u\,(3;\boldkey j)$ the generators are transformed as follows (see
[2]: $A_{01}=j_1A^*_{01}(\to)$, $A_{12}=j_2A^*_{12}(\to)$,
$A_{10}=j_1A^*_{10}(\to)$, $A_{21}=j_2A^*_{21}(\to)$,
$A_{kk}=A^*_{kk}(\to)$. Transformation of the components of
Gel'fand-Zetlin schemes can be defined as follows:
$$
\gathered
m_{13}=j_1j_2m^*_{13},\quad m_{23}=m^*_{23},\quad
m_{33}=j_1j_2m^*_{33},\\
m_{12}=j_2m^*_{12},\quad m_{22}=j_2m^*_{22},\quad m_{11}=m^*_{11}.
\endgathered
\tag 25
$$
Then the component of scheme $|\,m\rangle$ satisfy inequalities
$$
\spreadlines{1\jot}
\gathered
|\,m\rangle=\left|\left.\matrix
m_{13}\ m_{23}\ m_{33}\\
m_{12}\ m_{22}\\
m_{11}\endmatrix
\right.\right>,\\
{m_{13}\over j_1j_2}\geq m_{23}\geq{m_{33}\over j_1j_2},\quad
{m_{13}\over j_1j_2}\geq{m_{12}\over j_2}\geq m_{23},\\
m_{23}\geq{m_{22}\over j_2}\geq{m_{33}\over j_1j_2},\quad
{m_{12}\over j_2}\geq m_{11}\geq{m_{22}\over j_2}.
\endgathered
\tag 26
$$

Transforming the known expressions for generators of algebra $u\,(3)$
we come to generators of representations of algebra $u\,(3;\boldkey
j)$:
$$
\gather
A_{00}\,|\,m\rangle=\biggl(m_{23}+{m_{13}+m_{33}\over j_1j_2}
-{m_{12}+m_{22}\over j_2}\biggr)\,|\,m\rangle,\\ \vspace{1\jot}
A_{01}\,|\,m\rangle=\\ \vspace{1\jot}
={1\over j_2}\left\{-{\matrix
(m_{13}-j_1m_{12}+j_1j_2)(m_{33}-j_1m_{12}-j_1j_2)\times\\
\times(j_2m_{23}-m_{12})(j_2m_{11}-m_{12})
\endmatrix
\over(m_{22}-m_{12})(m_{22}-m_{12}-j_2)}\right\}^{1/2}
\,|\,m_{12}-j_2\rangle+\\ \vspace{1\jot}
+{1\over j_2}\left\{-{\matrix
(m_{13}-j_1m_{22}+2j_1j_2)(m_{33}-j_1m_{22})\times\\
\times(j_2m_{23}+j_2-m_{22})(j_2m_{11}+j_2-m_{22})
\endmatrix
\over(m_{12}-m_{22}+2j_2)(m_{12}-m_{22}+j_2)}\right\}^{1/2}
\,|\,m_{22}-j_2\rangle,\\ \vspace{1\jot}
A_{10}\,|\,m\rangle= \tag 27\\ \vspace{1\jot}
={1\over j_2}\left\{-{\matrix
(m_{13}-j_1m_{12})(m_{33}-j_1m_{12}-2j_1j_2)\times\\
\times(j_2m_{23}-j_2-m_{12})(j_2m_{11}-j_2-m_{12})
\endmatrix
\over(m_{22}-m_{12}-j_2)(m_{22}-m_{12}-2j_2)}\right\}^{1/2}
\,|\,m_{12}+j_2\rangle+\\  \vspace{1\jot}
+{1\over j_2}\left\{-{\matrix
(m_{13}-j_1m_{22}+j_1j_2)(m_{13}-j_1m_{22}-j_1j_2)\times\\
\times(j_2m_{23}-m_{22})(j_2m_{11}-m_{22})
\endmatrix
\over(m_{12}-m_{22}+j_2)(m_{12}-m_{22})}\right\}^{1/2}
\,|\,m_{22}+j_2\rangle,\\ \vspace{1\jot}
A_{02}\,|\,m\rangle=\\ \vspace{1\jot}
=\left\{-{\matrix
(m_{13}-j_1m_{12}+j_1j_2)(m_{33}-j_1m_{12}-j_1j_2)\times\\
\times(j_2m_{23}-m_{12})(m_{22}-j_2m_{11})
\endmatrix
\over(m_{22}-m_{12})(m_{22}-m_{12}-j_2)}\right\}^{1/2}
\left|\left.\matrix
m_{12}-j_2\\
m_{11}-1\endmatrix\right.\right>+
\endgather
$$
$$
\gathered
+\left\{-{\matrix
(m_{13}-j_1m_{22}+2j_1j_2)(m_{33}-j_1m_{22})\times\\
\times(j_2m_{23}+j_2-m_{22})(m_{12}-j_2m_{11}+j_2)
\endmatrix
\over(m_{12}-m_{22}+2j_2)(m_{12}-m_{22}+j_2)}\right\}^{1/2}
\left|\left.\matrix
m_{22}-j_2\\
m_{11}-1\endmatrix\right.\right>,\\ \vspace{1\jot}
A_{20}\,|\,m\rangle=\\ \vspace{1\jot}
=\left\{-{\matrix
(m_{13}-j_1m_{12})(m_{33}-j_1m_{22}-2j_1j_2)\times\\
\times(j_2m_{23}-j_2-m_{12})(m_{22}-j_2m_{11}-j_2)
\endmatrix
\over(m_{22}-m_{12}-j_2)(m_{22}-m_{12}-2j_2)}\right\}^{1/2}
\left|\left.\matrix
m_{12}+j_2\\
m_{11}+1\endmatrix\right.\right>+\\ \vspace{1\jot}
+\left\{-{\matrix
(m_{13}-j_1m_{22}+j_1j_2)(m_{33}-j_1m_{22}-j_1j_2)\times\\
\times(j_2m_{23}-m_{22})(m_{12}-j_2m_{11})
\endmatrix
\over(m_{12}-m_{22}+j_2)(m_{12}-m_{22})}\right\}^{1/2}
\left|\left.\matrix
m_{22}+j_2\\
m_{11}+1\endmatrix\right.\right>,
\endgathered
$$
where $|\,m_{12}\pm j_2\rangle$ is the scheme (26), in which
component $m_{12}$ is substituted for $m_{12}\pm j_2$ and so on.
Generators $A_{11}$, $A_{22}$, $A_{12}$, $A_{21}$, making subalgebra
$u(2;j_2)$,
are described by (7), where each index of generators must be
increased by unit and parameter $j_1$ has to be substituted for
parameter $j_2$.

Generators (27) satisfy the commutation relations of algebra
$u\,(3;\mathbreak\boldkey j)$:
$$
\gathered
[A_{kr},A_{pq}]={J_{kr}J_{rq}\over J_{kq}}\delta_{pr}A_{kq}-
{J_{kr}J_{rq}\over J_{pr}}\delta_{kq}A_{pr},\\
J_{kr}=\prod_{l=1+\min(k,r)}^{\max(k,r)}i_l,\quad
k,r,p,q=0,1,2,
\endgathered
\tag 28
$$
coinciding with Cartan-Weyl commutation
relations in [2] for $A_{kk}=H_k$, $A_{kr}=E_{l_k-l_r}$.

Unitary algebra $u\,(3)$ has three Casimir operators, which under
transition to algebra $u\,(3;\boldkey j)$ are transformed as follows
[2]:
$$
C_1(\boldkey j)=C^*_1(\to),\ C_2(\boldkey
j)=j_1^2j_2^2C^*_2(\to),\ C_3(\boldkey j)=j_1^2j_2^2C^*_3(\to).
$$
Spectrum of Casimir operators in this case
is as follows
$$
\gathered
C_1(\boldkey j)={m_{13}+m_{33}\over j_1j_2}+m_{23},\\
C_2(\boldkey j)=m^2_{13}+m^2_{33}+j_1^2j_2^2m^2_{23}+2j_1j_2
(m_{13}-m_{33}),\\
C_3(\boldkey j)={1\over j_1j_2}(m_{13}^3+m_{33}^3)+2(2m^2_{13}-
m_{33}^2)-m_{13}m_{33}+\\
+j_1^2j_2^2(m_{23}^3+2m_{23}^2-2m_{23})+\\
+j_1j_2[2(2m_{13}-m_{33})-m_{23}(m_{13}+m_{33})].
\endgathered
\tag 29
$$

This naturally brings up the question: for what considerations has
been chosen transformation rule (25) for components of
Gel'fand-Zetlin schemes or rule (6) in the case  of algebra
$u\,(2;j_1)$? We choose it in order to make spectrum of Casimir
operators of second order different from zero and not involving
undeterminate expressions for dual values of parameters $\boldkey j$.
Because $C_2(\boldkey j)=j_1^2j_2^2C^*_2(\to)$ and components
$m_{13}$, $m_{23}$, $m_{33}$ enter $C^*_2$ quadratically, this
requirement gives (25). However, variant (25) (we call it basic)
is not unique. Two other variants are possible as well:
$m_{13}=m^*_{13}$, $m_{23}=j_1j_2m^*_{23}$, $m_{33}=j_1j_2m^*_{33}$
or $m_{13}=j_1j_2m^*_{13}$, $m_{23}=j_1j_2m^*_{23}$,
$m_{33}=m^*_{33}$, which turn initial irreducible representation of
algebra $u\,(3)$ into representations of algebra $u\,(3;\boldkey j)$
with other (in comparison with basic invariant (29)) values of
Casimir operators. For example, $C'_2(\boldkey j)=m^2_{23}+m^2_{33}+
j_1^2j_2^2m_{13}(m_{13}+2)-2j_1j_2m_{33}$
and $C''_2(\boldkey j)=m_{13}^2+m^2_{23}+j_1^2j_2^2m_{33}(m_{33}-2)+
2j_1j_2m_{13}$.
The consideration of these variants of transition for irreducible
representations is quite similar to basic variant, and we skip the
corresponding relations.

It will be shown further that basic transformations (25), as two
other variants, give under contractions general, non-degenerate
representations of contracted algebras, all Casimir operators, which
are independent and have non-zero spectrum.

For interpretation of formal inequalities (26) let us consider the action of
rising generator $A_{10}$ on the vector of the ``major weight''
$\varphi_{\scriptscriptstyle\text{Mw}}$, described by scheme (2.26),
for $m_{11}=m_{12}=m_{13}$, $m_{22}=m_{23}$, and the action of
lowering generator $A_{01}$ on the vector of the ``minor weight''
$\varphi_{\scriptscriptstyle\text{mw}}$,
described by scheme (26) for $m_{11}=m_{22}=m_{33}$,
$m_{12}=m_{23}$. Let us write out explicitly  only those factors,
which vanish for $j_1=j_2=1$. Then
$$
\gathered
A_{10}\varphi_{\scriptscriptstyle\text{Mw}}={1\over
j_2}\{m_{13}(1-j_1)A\}^{1/2}\,|\,m_{13}+j_2\rangle+\\
+{1\over j_2}\{m_{23}(j_2-1)B\}^{1/2}\,|\,m_{23}+j_2\rangle,\\
A_{01}\varphi_{\scriptscriptstyle\text{mw}}={1\over
j_2}\{m_{23}(j_2-1)C\}^{1/2}\,|\,m_{23}-j_2\rangle+\\
+{1\over j_2}\{m_{33}(1-j_1)D\}^{1/2}\,|\,m_{33}-j_2\rangle.
\endgathered
\tag 30
$$
It can be seen from here that for $j_1\neq1$, $j_2=1$
$A_{10}\varphi_{\scriptscriptstyle\text{Mw}}=\{m_{13}(1-\mathbreak
-j_1)A\}^{1/2}
\,|\,m_{13}+1\rangle\neq0$,
which means the absence of the bound from above on $m_{12}$;
$A_{01}\varphi_{\scriptscriptstyle\text{mw}}=\{m_{33}(1-j_1)D\}^{1/2}
\,|\,m_{33}-1\rangle\neq0$,
that means the absence of the bound from below on $m_{22}$, i.e.
components of scheme (26) satisfy inequalities $m_{12}\geq
m_{23}\geq m_{22}$. For $j_1=1$, $j_2\neq1$ we obtain from (30)
$A_{10}\varphi_{\scriptscriptstyle\text{Mw}}=
{1\over j_2}\{m_{23}(j_2-1)B\}^{1/2}\,|\,m_{23}+j_2\rangle\neq0$,
that means the absence of the bound from above on $m_{22}$, and
$A_{01}\varphi_{\scriptscriptstyle\text{mw}}={1\over
j_2}\{m_{23}(j_2-1)C\}^{1/2}\,|\,m_{23}-j_2\rangle\neq0$,
which means the absence of the bound from below on $m_{12}$, i.e.
components of scheme (26) satisfy inequalities $m_{13}\geq
\mathbreak\geq m_{12}$, $m_{22}\geq m_{33}$, $-\infty<m_{11}<\infty$.
At last, we find from (30) for $j_1\neq1$, $j_2\neq1$ that there are
no restrictions for components $m_{12}$, $m_{22}$, $m_{11}$. In all
cases inequality $m_{13}\geq m_{33}$ remains valid.

The same inequalities for components of Gel'fand-Zetlin scheme can be
derived from formal inequalities (26), if one interprets them for
$j_1,\mathbreak j_2\neq1$ according to following rules: inequality
${m\over j}\geq m_1$ means the absence of the bounds from above on
$m_1$; inequality $m_1\geq{m\over j}$ means the absence of the bounds
from below on $m_1$; inequality ${m\over j_1j_2}\geq{m_1\over j_2}$ is
equivalent to ${m\over j_1}\geq m_1$, i.e. common parameters in both
parts of inequality can be cancelled out. The same rules are valid for
algebras of higher dimensions as well.

Formulas for irreducible representations of algebra $u\,(3)$ can be
obtained from formulas of this paragraph for $j_1=j_2=1$. The
requirement of unitarity for representations of algebra $u\,(3)$ leads
to the following relations for operators (27):
$A_{kk}=\bar{A}_{kk}$ $(k=0,1,2)$, $A_{rp}=\bar{A}_{pr}$
$(r,p=0,1,2)$. Here the bar means complex conjugation.

\vfill\eject
{\bf 2.2. Contraction over the first parameter}
\medskip

The structure of contracted unitary algebra, described in [2], is
as follows: $u\,(3;\iota_1,j_2)=T_4\notni(u\,(1)\oplus u\,(2;j_2))$,
where $T_4=\{A_{01},A_{10},A_{02},\mathbreak A_{20}\}$;
$u\,(2;j_2)=\{A_{11}, A_{22},A_{12},A_{21}\}$; $u\,(1)=\{A_{00}\}$.
The relations (27) give for $j_1=\iota_1$:
$$
\gathered
A_{00}\,|\,m\rangle=\biggl({m_{13}+m_{33}\over\iota_1j_2}
+m_{23}-{m_{12}+m_{22}\over j_2}\biggr)\,|\,m\rangle,\\
A_{01}\,|\,m\rangle={1\over j_2}\sqrt{-m_{13}m_{33}}\times\\
\times\biggl(\left\{{(j_2m_{23}-m_{12})(j_2m_{11}-m_{12})\over
(m_{22}-m_{12})(m_{22}-m_{12}-j_2)}\right\}^{1/2}\,|\,m_{12}-j_2\rangle+\\
+\left\{{(j_2m_{23}+j_2-m_{22})(j_2m_{11}+j_2-m_{22})\over
(m_{12}-m_{22}+2j_2)(m_{12}-m_{22}+j_2)}\right\}^{1/2}\,|\,m_{22}-j_2\rangle
\biggr),\\
A_{10}\,|\,m\rangle={1\over j_2}\sqrt{-m_{13}m_{33}}\times\\
\times\biggl(\left\{{(j_2m_{23}-j_2-m_{12})(j_2m_{11}-j_2-m_{12})\over
(m_{22}-m_{12}-j_2)(m_{22}-m_{12}-2j_2)}\right\}^{1/2}\,|\,m_{12}+j_2\rangle+\\
+\left\{{(j_2m_{23}-m_{22})(j_2m_{11}-m_{22})\over
(m_{12}-m_{22}+j_2)(m_{12}-m_{22})}\right\}^{1/2}\,|\,m_{22}+j_2\rangle
\biggr),\\
A_{02}\,|\,m\rangle=\sqrt{-m_{13}m_{33}}\times\\
\times\biggl(\left\{{(j_2m_{23}-m_{12})(m_{22}-j_2m_{11})\over
(m_{22}-m_{12})(m_{22}-m_{12}-j_2)}\right\}^{1/2}\left|\left.\matrix
m_{12}-j_2\\
m_{11}-1\endmatrix\right.\right>+\\
+\left\{{(j_2m_{23}+j_2-m_{22})(m_{12}-j_2m_{11}+j_2)\over
(m_{12}-m_{22}+2j_2)(m_{12}-m_{22}+j_2)}\right\}^{1/2}\left|\left.\matrix
m_{22}-j_2\\
m_{11}-1\endmatrix\right.\right>
\biggr),\\
A_{20}\,|\,m\rangle=\sqrt{-m_{13}m_{33}}\times\\
\times\biggl(\left\{{(j_2m_{23}-j_2-m_{12})(m_{22}-j_2m_{11}-j_2)\over
(m_{22}-m_{12}-j_2)(m_{22}-m_{12}-2j_2)}\right\}^{1/2}\left|\left.\matrix
m_{12}-j_2\\
m_{11}+1\endmatrix\right.\right>+\\
+\left\{{(j_2m_{23}-m_{22})(m_{12}-j_2m_{11})\over
(m_{12}-m_{22}+j_2)(m_{12}-m_{22})}\right\}^{1/2}\left|\left.\matrix
m_{22}+j_2\\
m_{11}+1\endmatrix\right.\right>
\biggr).
\endgathered
\tag 31
$$
Here dual parts, arising in the expressions
for generators, are omitted, and only real parts are written.

Algebra $u\,(3;\iota_1,1)$ is inhomogeneous algebra $iu(2)$ in
Chakrabarti's notations [7]. The requirement of determinacy and
unitarity of generator $A_{00}$ gives
$(m_{13}+m_{33})/\iota_1=\xi\in\Bbb R$, i.e.
$$
m_{13}=k+\iota_1\xi/2,\
m_{33}=-k+\iota_1\xi/2,\quad \xi,k\in\Bbb R,\quad k\geq0.
\tag32
$$
Real-valuedness of $k$ follows from unitary relations for $A_{01}$,
$A_{10}$, and its positiveness -- from inequality $m_{13}\geq m_{33}$,
considered for real parts. Taking into account (32), we get
$\sqrt{-m_{13}m_{33}}=k$, and the expressions (2.31) for
$j_2=1$ coincide with corresponding Chakrabarti's formulas [7] for
$iu(2)$.  The integer components of scheme $|\,\widetilde{m}\rangle$
are interrelated via inequalities $m_{12}\geq m_{23}\geq m_{22}$,
$m_{12}\geq m_{11}\geq m_{22}$, ensued from (26) for $j_1=\iota_1$.
The scheme $|\,\widetilde{m}\rangle$ can be obtained from scheme
(26) for $m_{13}=k$, $m_{33}=-k$. Spectrum of Casimir operators in a
given irreducible representation of algebra $u\,(3;\iota_1,1)$ can be
found from (29):
$$
\gathered
C_1(\iota_1,1)=\xi+m_{23},\quad C_2(\iota_1,1)=2k^2,\\
C_3(\iota_1,1)=3k^2(\xi+1).
\endgathered
\tag 33
$$
Algebra $su\,(3;\iota_1,1)$ differs from algebra $u\,(3;\iota_1,1)$ in
that diagonal operators satisfy the relation $A_{00}+A_{11}+A_{22}=0$.
Acting on scheme $|\,\widetilde{m}\rangle$, we get
$\xi+m_{23}\,|\,\widetilde{m}\rangle=0$, from which it follows
$\xi=-m_{23}$.  Substituting $\xi$ in (33), we find spectrum of
Casimir operators
$$
C_1(\iota_1,1)=0,\quad C_2(\iota_1,1)=2k^2,\quad
C_3(\iota_1,1)=3k^2(1-m_{23})
\tag 34
$$
of the irreducible representation of
algebra $su\,(3;\iota_1,1)=T_4\notni u\,(2)$, generators of which are
described by (31) for $j_2=1$, where it is necessary to put
$$
m_{13}=k-\iota_1m_{23}/2,\quad m_{33}=-k-\iota_1m_{23}/2,\quad
k\geq0,\quad m_{23}\in\Bbb Z.
\tag 35
$$
\pagebreak Here $\Bbb Z$ is a set of integers.

{\bf 2.3. Contraction over the second parameter}
\medskip

The structure of contracted algebra is described in [2] and is as
follows:  $u\,(3;j_1,\iota_2)=T_4\notni(u\,(2;j_1)\oplus u\,(1))$,
where $T_4=\{A_{12},A_{21},A_{02}, A_{20}\}$;
$u\,(2;j_1)=\{A_{00}, A_{11},A_{01},A_{10}\}$; $u\,(1)=\{A_{22}\}$.
After substitution of $j_2=\iota_2$ in (27), expressions
$|\,m_{12}\pm\iota_2\rangle$ can occur, with  which we proceed
according to  general rules (see [6]) of treating functions
of  dual variable,  i.e.  expand into series
$$
\gathered
|\,m_{12}\pm\iota_2\rangle=|\,m\rangle\pm\iota_2\,|\,m\rangle'_{12},
\quad|\,m\rangle'_{12}\equiv{\partial\over\partial m_{12}}\,|\,m\rangle,\\
|\,m_{22}\pm\iota_2\rangle=|\,m\rangle\pm\iota_2\,|\,m\rangle'_{22},
\quad|\,m\rangle'_{22}\equiv{\partial\over\partial m_{22}}\,|\,m\rangle.
\endgathered
\tag 36
$$
Taking this remark into account, (27) give  for $j_2=\iota_2$
the following expressions for generators:
$$
\gathered
A_{00}\,|\,m\rangle=\biggl(m_{23}+{m_{13}+m_{33}\over\iota_2j_1}-
{m_{12}+m_{22}\over\iota_2}\biggr)\,|\,m\rangle,\\
A_{11}\,|\,m\rangle=\biggl({m_{12}+m_{22}\over\iota_2}-m_{11}\biggr)
\,|\,m\rangle,\quad A_{22}\,|\,m\rangle=m_{11}\,|\,m\rangle,\\
A_{12}\,|\,m\rangle=\sqrt{-m_{12}m_{22}}\,|\,m_{11}-1\rangle,\\
A_{21}\,|\,m\rangle=\sqrt{-m_{12}m_{22}}\,|\,m_{11}+1\rangle,\\
A_{01}\,|\,m\rangle={1\over m_{12}-m_{22}}\left\{{1\over\iota_2}
(m_{12}\alpha_{12}+m_{22}\alpha_{22})\,|\,m\rangle+\right.\\
+{1\over2\alpha_{12}}\left[j_1m_{12}(m_{13}-m_{33})-\right.\\
\left.-\alpha_{12}^2\biggl(m_{11}+m_{23}+{m_{12}\over
m_{12}-m_{22}}\biggr) \right]\,|\,m\rangle-\\
-{1\over2\alpha_{22}}\left[2j_1m_{22}(m_{33}-j_1m_{22})+\right.\\
\left.+\alpha_{22}^2\biggl(m_{11}+m_{23}+{2m_{12}+m_{22}\over
m_{12}-m_{22}}
\biggr)\right]\,|\,m\rangle-m_{12}\alpha_{12}\,|\,m\rangle'_{12}-\\
-\left.m_{22}\alpha_{22}\,|\,m\rangle'_{22}\right\},\\
A_{10}\,|\,m\rangle={1\over m_{12}-m_{22}}\left\{{1\over\iota_2}
(m_{12}\alpha_{12}+m_{22}\alpha_{22})\,|\,m\rangle+\right.
\endgathered
$$
$$
\gathered
+{1\over2\alpha_{12}}\left[2j_1m_{12}(m_{13}-j_1m_{12})-\right.\\
\left.-\alpha_{12}^2\biggl(m_{11}+m_{23}+{m_{12}+2m_{22}\over
m_{12}-m_{22}}\biggr)\right]\,|\,m\rangle+\\
+{1\over2\alpha_{22}}\left[j_1m_{22}(m_{13}-m_{33})-\right.\\
\left.-\alpha_{22}\biggl(m_{11}+m_{23}+{m_{22}\over m_{12}-m_{22}}
\biggr)\right]\,|\,m\rangle+m_{12}\alpha_{12}\,|\,m\rangle'_{12}+\\
\left.+m_{22}\alpha_{22}\,|\,m\rangle'_{22}\right\},\\
A_{02}\,|\,m\rangle={\sqrt{-m_{12}m_{22}}\over m_{12}-m_{22}}
(\alpha_{12}+\alpha_{22})\,|\,m_{11}-1\rangle,\\
A_{20}\,|\,m\rangle={\sqrt{-m_{12}m_{22}}\over m_{12}-m_{22}}
(\alpha_{12}+\alpha_{22})\,|\,m_{11}+1\rangle,\\
\alpha_{12}=\sqrt{-(m_{13}-j_1m_{12})(m_{33}-j_1m_{12})},\\
\alpha_{22}=\sqrt{-(m_{13}-j_1m_{22})(m_{33}-j_1m_{22})},
\endgathered
\tag 37
$$
where only real parts are written. To obtain correct expressions for
$A_{01}$, $A_{10}$, it is necessary to consider in denominators of
(27) $m_{12}-m_{22}$.

The requirement of determinacy of generators $A_{00}$, $A_{11}$
together with condition of their Hermiticity brings to components of
schemes
$$
\gathered
m_{13}=k+\iota_2j_1\xi/2,\quad m_{33}=-k+\iota_2j_1\xi/2,\quad
k\geq0, \xi\in\Bbb R,\\
m_{12}=r+\iota_2\zeta/2,\quad m_{22}=-r+\iota_2\zeta/2,\quad
r\geq0,\quad \zeta\in\Bbb R.
\endgathered
\tag 38
$$
Substituting components (38) in (26), we get
$$
\gathered
|\,m\rangle=|\,\widetilde{m}\rangle+\iota_2j_1{\xi\over2}
(|\,\widetilde{m}\rangle'_{13}+|\,\widetilde{m}\rangle'_{33})+
\iota_2{\zeta\over2}
(|\,\widetilde{m}\rangle'_{12}+|\,\widetilde{m}\rangle'_{22}),\\
|\,\widetilde{m}\rangle=\left|\left.\matrix
k\ \ m_{23}\ \ -k\\
r\ \ -r\\
m_{11}\endmatrix\right.\right>,\quad m_{11},m_{23}\in\Bbb Z.
\endgathered
\tag 39
$$

For classical unitary algebras, Gel'fand-Zetlin schemes $|\,m\rangle$
with integer components enumerate normalized to unit basis vectors in
finite-dimensional space of representation. Under contraction and
analytical continuations a part of components of schemes
$|\,m\rangle$ takes continuous values. In this case basis vectors in
infinite-dimensional space of representation for contracted or
analytically continued algebras, corresponding to such schemes are
understood to be generalized functions, orthogonal as before, but
normalized to delta- function. In particular, for
$|\,\widetilde{m}\rangle$ we get
$$
\langle\widetilde{m}'\,|\,\widetilde{m}\rangle=\delta^2(k'-k)
\delta^2(r'-r)\delta_{m'_{23},m_{23}}\delta_{m'_{11},m_{11}},
\tag 40
$$
where the squared delta-functions
occur due to the fact that $r$ and $k$ twice enter the components of
scheme (for details see [12--14] by Celeghini in the case of
contractions and [15, 16] in the case of analytical
continuations).

Substituting (38), (39) in (37), we obtain generators of
irreducible representation of algebra $u\,(3;j_1,\iota_2)$:
$$
\gathered
A_{00}\,|\,\widetilde{m}\rangle=(m_{23}+\xi-\zeta)\,|\,
\widetilde{m}\rangle,\quad A_{11}\,|\,\widetilde{m}\rangle=
(\zeta-m_{11})\,|\,\widetilde{m}\rangle,\\
A_{22}\,|\,\widetilde{m}\rangle=m_{11}\,|\,\widetilde{m}\rangle,\\
A_{12}\,|\,\widetilde{m}\rangle=r\,|\,\widetilde{m_{11}-1}\rangle,
\quad
A_{11}\,|\,\widetilde{m}\rangle=r\,|\,\widetilde{m_{11}+1}\rangle,\\
A_{02}\,|\,\widetilde{m}\rangle=\sqrt{k^2-j_1^2r^2}\,|\,
\widetilde{m_{11}-1}\rangle,\\
A_{20}\,|\,\widetilde{m}\rangle=\sqrt{k^2-j_1^2r^2}\,|\,
\widetilde{m_{11}+1}\rangle,\\
A_{01}\,|\,\widetilde{m}\rangle={1\over2r}\sqrt{k^2-j_1^2r^2}
\left\{\biggl(\zeta-m_{11}-m_{23}-{1\over2}\biggr)\right.\,|\,
\widetilde{m}\rangle+\\
+\left.j_1^2r^2{\xi-\zeta+1\over
k^2-j_1^2r^2}\,|\,\widetilde{m}\rangle-
r(|\,\widetilde{m}\rangle'_{12}-|\,\widetilde{m}\rangle'_{22})\right\},\\
A_{10}\,|\,\widetilde{m}\rangle={1\over2r}\sqrt{k^2-j_1^2r^2}
\left\{\biggl(\zeta-m_{11}-m_{23}+{1\over2}\biggr)\right.\,|\,
\widetilde{m}\rangle+\\
+\left.j_1^2r^2{\xi-\zeta-1\over
k^2-j_1^2r^2}\,|\,\widetilde{m}\rangle+r(|\,\widetilde{m}\rangle'_{12}
-|\,\widetilde{m}\rangle'_{22})\right\}.
\endgathered
\tag 41
$$
The relation of Hermiticity for operators $A_{02}$, $A_{20}$ gives
$k^2-j_1^2r^2\geq0$, which for $j_1=1$ imposes restriction $k\geq r$.
The action of operators on the derived schemes can be found, using
(36), by application of operators to both sides of equation
$|\,m\rangle'_{12}={1\over2\iota_2}(|\,m_{12}+\iota_2\rangle-
|\,m_{12}-\iota_2\rangle)$. The eigenvalues of Casimir operators for
representation (41) can be obtained by substituting of components
(38) in (29). They are as follows:
$$
\gathered
C_1(j_1,\iota_2)=\xi+m_{23},\quad C_2(j_1,\iota_2)=2k^2,\\
C_3(j_1,\iota_2)=3k^2(\xi+1).
\endgathered
\tag 42
$$
They are all different from
zero and independent, as it must be for non-degenerate irreducible
representations of algebra $u\,(3;j_1,\iota_2)$. Let us notice that
spectrum (42) coincides with spectrum (33) of Casimir operators
for algebra $u\,(3;\iota_1,j_2)$.

For the sake of convenience of applications (interpretation) we have
fixed indices of generators $A_{ps}$, for this reason
$u\,(3;1,\iota_2)$ and $u\,(3;\iota_1,1)$ turned out in our case to be
different algebras. Rejecting this agreement it is easy to prove that
these algebras are isomorphic. Representation (31), (32) is
realized in discrete basis, generated by the chain of subalgebras
$u\,(3;\iota_1,1)$ $\supset$ $u\,(2;1)$ $\supset$ $u\,(1)$ and
described by schemes:
$$
\left|\left.\matrix
k\ \ m_{23}\ \ -k\\
m_{12}\ \ m_{22}\\
m_{11}\endmatrix\right.\right>,\quad
\matrix\format\l\\
m_{23}\in\Bbb Z,\ \ k\geq0,\\
m_{12}\geq m_{23}\geq m_{22},
\ \ m_{12}\geq m_{11}\geq m_{22},\\
m_{12},m_{22},m_{11}\in\Bbb Z,\endmatrix
\tag 43
$$
whereas representation (41) is realized in
continuous basis, generated by expansion $u\,(3;1,\iota_2)$ $\supset$
$u\,(2;\iota_2)$ $\supset$ $u\,(1)$ and described by schemes
$$
\left|\left.\matrix
k\ \ m_{23}\ \ -k\\
r\ \ -r\\
m_{11}\endmatrix\right.\right>,\quad
\matrix\format\l\\
m_{23},m_{11}\in\Bbb Z,\\
k\geq r\geq0,\endmatrix
\tag 44
$$
where besides $A_{00}$, $A_{11}$, $A_{22}$ operator $A_{01}+A_{10}$ is
also diagonal in this basis.

Thus, contractions over different parameters, leading to isomorphic
algebras, give description of the same irreducible representation of
contracted algebra in different basises, generated by canonical chains
of subalgebras.

\vfill\eject

{\bf 2.4. Two-dimensional contraction}
\medskip

The  structure  of  algebra  $u\,(3;\boldsymbol\iota)$  is  given
in [2]. It is as follows:
$u\,(3;\boldsymbol\iota)=T_6\notni(\{A_{00}\}\oplus\{A_{11}\}
\oplus\{A_{22}\})$, where nilpotent subalgebra $T_6$ is spanned over
generators $A_{ps}$, $p,s=1,2$. The explicit form of generators of
irreducible representations  of algebra can be obtained, putting
either $j_1=\iota_1$, $j_2=\iota_2$ in (27), or $j_2=\iota_2$ in
(31), or from (37) for $j_1=\iota_1$. All  three approaches lead
to the same  result:
$$
\gather
A_{00}\,|\,m\rangle=\biggl(m_{23}+{m_{13}+m_{33}\over\iota_1\iota_2}-
{m_{12}+m_{22}\over\iota_2}\biggr)\,|\,m\rangle,\\
A_{22}\,|\,m\rangle=m_{11}\,|\,m\rangle,\\
A_{11}\,|\,m\rangle=\biggl({m_{12}+m_{22}\over\iota_2}-m_{11}\biggr)
\,|\,m\rangle,\quad A_{12}\,|\,m\rangle=\alpha\,|\,m_{11}-1\rangle,\\
A_{21}\,|\,m\rangle=\alpha\,|\,m_{11}+1\rangle,\\
A_{02}\,|\,m\rangle={2\alpha\beta\over
m_{12}-m_{22}}\,|\,m_{11}-1\rangle, \quad
A_{20}\,|\,m\rangle={2\alpha\beta\over
m_{12}-m_{22}}\,|\,m_{11}+1\rangle,\\
A_{01}\,|\,m\rangle={\beta\over
m_{12}-m_{22}}\left\{{m_{12}+m_{22}\over
\iota_2}\,|\,m\rangle-\right. \tag 45\\
\left.-\biggl(m_{11}+m_{23}+{3m_{12}+m_{22}\over2(m_{12}-m_{22})}\biggr)
\,|\,m\rangle-m_{12}\,|\,m\rangle'_{12}-m_{22}\,|\,m\rangle'_{22}\right\},\\
A_{10}\,|\,m\rangle={\beta\over
m_{12}-m_{22}}\left\{{m_{12}+m_{22}\over
\iota_2}\,|\,m\rangle-\right.\\
\left.-\biggl(m_{11}+m_{23}+{m_{12}+3m_{22}\over2(m_{12}-m_{22})}\biggr)
\,|\,m\rangle+m_{12}\,|\,m\rangle'_{12}+m_{22}\,|\,m\rangle'_{22}\right\},\\
\alpha=\sqrt{-m_{12}m_{22}},\quad\beta=\sqrt{-m_{13}m_{33}}.
\endgather
$$
The  requirement of  determinacy of  operators $A_{00}$, $A_{11}$  and
condition  of Hermiticity,   give   for   the   components   of
scheme   $|\,m\rangle$:
$$
\gathered
m_{13}=k+\iota_1\iota_2\xi/2,\quad m_{33}=-k+\iota_1\iota_2\xi/2,
\quad k\geq0,\quad\xi\in\Bbb R,\\
m_{12}=r+\iota_2\zeta/2,\quad m_{22}=-r+\iota_2\zeta/2,
\quad r\geq0,\quad\zeta\in\Bbb R.
\endgathered
\tag 46
$$
The   substitution   of   these
expressions in (45) leads to representation operators
$$
\gathered
A_{00}\,|\,\widetilde{m}\rangle=
(m_{23}+\xi-\zeta)\,|\,\widetilde{m}\rangle,\quad
A_{11}\,|\,\widetilde{m}\rangle=(\zeta-m_{11})\,|\,\widetilde{m}\rangle,\\
A_{22}\,|\,\widetilde{m}\rangle=m_{11}\,|\,\widetilde{m}\rangle,
\quad
A_{12}\,|\,\widetilde{m}\rangle=r\,|\,\widetilde{m_{11}-1}\rangle,\\
A_{21}\,|\,\widetilde{m}\rangle=r\,|\,\widetilde{m_{11}+1}\rangle,\\
A_{02}\,|\,\widetilde{m}\rangle=k\,|\,\widetilde{m_{11}-1}\rangle,
\quad
A_{20}\,|\,\widetilde{m}\rangle=k\,|\,\widetilde{m_{11}+1}\rangle,\\
A_{01}\,|\,\widetilde{m}\rangle={k\over2r}\biggl(\zeta-m_{11}-m_{23}
-{1\over2}\biggr)\,|\,\widetilde{m}\rangle-\\
-{k\over2}(|\,\widetilde{m}\rangle'_{12}-|\,\widetilde{m}\rangle'_{22}),\\
A_{10}\,|\,\widetilde{m}\rangle={k\over2r}\biggl(\zeta-m_{11}-m_{23}
+{1\over2}\biggr)\,|\,\widetilde{m}\rangle+\\
+{k\over2}(|\,\widetilde{m}\rangle'_{12}-|\,\widetilde{m}\rangle'_{22}),
\endgathered
\tag 47
$$
where $|\,\widetilde{m}\rangle$ means the scheme
$$
|\,\widetilde{m}\rangle=
\left|\left.\matrix
k\ \ m_{23}\ \ -k\\
r\ \ -r\\
m_{11}\endmatrix\right.\right>,\quad
\matrix\format\l\\
m_{11},m_{23}\in\Bbb Z,\\
k\geq 0,\ r\geq0.\endmatrix
\tag 48
$$
It is worth of  mentioning that operator $A_{01}+A_{10}$  is diagonal
in basis  $|\,\widetilde{m}\rangle$, and  spectrum  of  Casimir
operators  for irreducible representations (47) of algebra
$u\,(3;\boldsymbol\iota)$ is given  by the same formulas (33),
(42)  as in the case of algebras $u\,(3;\iota_1,j_2)$,
$u\,(3;j_1,\iota_2)$.
\bigskip

{\bf 3. Representations of unitary algebras $u\,(n;\boldkey j)$}
\medskip

{\bf 3.1. Operators of representation}
\medskip

Standard notations of Gel'fand-Zetlin [11] correspond to diminishing
chain of subalgebras $u\,(n)$ $\supset$ $u\,(n-1)$ $\supset$ $\dots$
$\supset$ $u\,(2)$ $\supset$ $u\,(1)$, where
$u\,(n)=\{E_{kr},\ k,r=1,2,\dots,n\}$,
$u\,(n-1)=\{E_{kr},\ k,r=1,2,\dots,n-\mathbreak-1\},\,\dots\,,$
$u\,(2)=\{E_{kr},\ k,r=1,2\}$, $u\,(1)=\{E_{11}\}$. We shall use
now another imbedding of subalgebra into algebra, which leads to the
chain of subalgebras $u\,(n;j_1,j_2,\dots,j_{n-1})$ $\supset$
$u\,(n-1;j_2,\dots,j_{n-1})$
$\supset\,\dots\,\supset\mathbreak\supset$ $u\,(2;j_{n-1})$ $\supset$
$u\,(1)$, where $u\,(n;j_1,\dots,j_{n-1})=\{A_{sp},
\ p,s=0,1,\dots,\mathbreak n-1\}$,
$u\,(n-1;j_2,\dots,j_{n-1})=\{A_{sp},\ s,p=1,2,\dots,n-1\}$,
$\dots,$ $u\,(2;\mathbreak j_{n-1})=\{A_{sp},\ s,p=n-2,n-1\}$,
$u\,(1)=\{A_{n-1,n-1}\}$. To pass from standard notations to ours, it
is necessary to change index $k$ of generator for index $n-k$ and to
leave unchanged the numbering of components in Gel'fand-Zetlin
schemes.

To determine representations of algebra $u\,(n)$, it is sufficient to
give the action of generators $E_{kk}$, $E_{k,k+1}$, $E_{k+1,k}$ and
to find the rest generators from commutators. In our notations it is
sufficient to know generators $A_{n-k,n-k}$, $A_{n-k,n-k-1}$,
$A_{n-k-1,n-k}$, which are transformed under transition from $u\,(n)$
to $u\,(n;\boldkey j)$ as follows:
$$
\gathered
A_{n-k,n-k-1}=j_{n-k}A^*_{n-k,n-k-1}(\to),\\
A_{n-k-1,n-k}=j_{n-k}A^*_{n-k-1,n-k}(\to),\quad k=1,2,\dots,n-1,
\endgathered
\tag 49
$$
where $j_{n-k}$ for dual value plays
the role of tending to zero parameter in Wigner-Inenu contraction
[17]; $A^*(\to)$ is singularly transformed generator.

To give of a singular transformation is equivalent as to give the
transformation rule for components of Gel'fand-Zetlin scheme
$$
\gathered
|\,m^*\rangle=\left|\left.\matrix
m^*_{1n}\ \ \ \ \ \ m^*_{2n}\,\dots\,m^*_{n-1,n}\ \ \ \ \ \ m^*_{nn}\\
m^*_{1,n-1}\ \ m^*_{2,n-1}\,\dots\,m^*_{n-1,n-1}\\
\dots\dots\dots\dots\\
m^*_{12}\ \ m^*_{22}\\
m^*_{11}\endmatrix\right.\ \ \right\rangle,\\
m^*_{pk}\geq m^*_{p,k-1}\geq m^*_{p+1,k},\\
k=2,3,\dots,n,\\
p=1,2,\dots,n-1,\\
m^*_{1n}\geq m^*_{2n}\geq\dots\geq m^*_{nn}
\endgathered
\tag 50
$$
under the transition from $u\,(n)$ to $u\,(n;\boldkey j)$. Defining
this transformation by
$$
\gathered
m_{1k}=m^*_{1k}J_k,\quad m_{kk}=m^*_{kk}J_k,\quad
J_k=\prod_{l=n-k+1}^{n-1}j_l,\\
m_{pk}=m^*_{pk},\quad p=2,3,\dots,n-1,\quad k=2,3,\dots,n,
\endgathered
\tag 51
$$
we obtain the scheme $|\,m\rangle$, which
components $m_{pk}$ are integers, and components $m_{1k}$, $m_{kk}$
can be complex or dual numbers. They satisfy inequalities
$$
\gather
m_{pk}\geq m_{p,k-1}\geq m_{p+1,k},\quad k=2,3,\dots,n,\quad
p=2,3,\dots,n-2,\\
{m_{1k}\over J_k}\geq{m_{1,k-1}\over J_{k-1}}\geq m_{2k},\quad
m_{k-1,k}\geq{m_{k-1,k-1}\over J_{k-1}}\geq{m_{kk}\over J_k},
\tag 52\\
{m_{1n}\over J_n}\geq m_{2n}\geq m_{3n}\geq\dots\geq m_{n-1,n}\geq
{m_{nn}\over J_n},
\endgather
$$
which for dual and imaginary values
of parameters $\boldkey j$ are interpreted according to the rules,
described in $\S$2.1.

Substituting (51) in known expressions for generators of algebra
and taking into account (49), we find operators of representation of
algebra $u\,(n;\boldkey j)$:
$$
\gather
A_{n-k,n-k}\,|\,m\rangle=\biggl({m_{1k}+m_{kk}\over J_k}-
{m_{1,k-1}+m_{k-1,k-1}\over J_{k-1}}+m_{k-1,k}+\\
+\sum_{s=2}^{k-2}(m_{sk}-m_{s,k-1})\biggr)\,|\,m\rangle,\quad
k=1,2,\dots,n,\\
A_{n-k-1,n-k}\,|\,m\rangle={1\over J_k}[\widetilde{a}^1_k(m)\,|\,
m_{1k}-J_k\rangle+\widetilde{a}^k_k(m)\,|\,m_{kk}-J_k\rangle]+\\
+j_{n-k+1}\sum_{s=2}^{k-1}\widetilde{a}^s_k(m)\,|\,m_{sk}-1\rangle,
\tag 53\\
A_{n-k,n-k-1}\,|\,m\rangle={1\over J_k}[\widetilde{b}^1_k(m)\,|\,
m_{1k}+J_k\rangle+\widetilde{b}^k_k(m)\,|\,m_{kk}+J_k\rangle]+\\
+j_{n-k+1}\sum_{s=2}^{k-1}\widetilde{b}^s_k(m)\,|\,m_{sk}+1\rangle,
\quad k=1,2,\dots,n-1,\\
\endgather
$$
where
$$
\gather
\widetilde{a}^1_k(m)=\left\{{\prod\limits_{p=2}^k(J_kl_{p,k+1}-l_{1k}+J_k)
\prod\limits_{p=2}^{k-2}(J_kl_{p,k-1}-l_{1k})\over\prod\limits_{p=2}^{k-1}
(J_kl_{pk}-l_{1k}+J_k)(J_kl_{pk}-l_{1k})}\right\}^{{1\over2}}\times\\
\times\left\{-{\matrix
(l_{1,k+1}-j_{n-k}l_{1k}+J_{k+1})(l_{k+1,k+1}-j_{n-k}l_{1k}+J_{k+1})
\times\\
\times(l_{1,k-1}j_{n-k+1}-l_{1k})\endmatrix
\over(l_{kk}-l_{1k}+J_k)(l_{kk}-l_{1k})(l_{k-1,k-1}j_{n-k+1}-l_{1k})^{-1}}
\right\}^{{1\over2}},\\
\widetilde{a}^s_k(m)=\left\{{\prod\limits_{p=2}^k(l_{p,k+1}-l_{sk}+1)
\prod\limits_{p=2}^{k-2}(l_{p,k-1}-l_{sk})\over\prod\limits^{k-1}\Sb
p=2\\ p\neq s\endSb
(l_{pk}-l_{sk}+1)(l_{pk}-l_{sk})}\right\}^{{1\over2}}\times\\
\times\left\{-{(l_{1,k+1}-J_{k+1}l_{sk}+J_{k+1})\over
(l_{1k}-J_kl_{sk}+J_k)}\right.\times \tag 54\\
\times\left.{\matrix
(l_{k+1,k+1}-J_{k+1}l_{sk}+J_{k+1})(l_{1,k-1}-J_{k-1}l_{sk})
\times\\
\times(l_{k-1,k-1}-J_{k-1}l_{sk})\endmatrix
\over(l_{1k}-J_kl_{sk})(l_{kk}-J_kl_{sk}+J_k)(l_{kk}-J_kl_{sk})}
\right\}^{{1\over2}},\quad 1<s<k,\\
\widetilde{b}^1_k(m)=\left\{{\prod\limits_{p=2}^k(J_kl_{p,k+1}-l_{1k})
\prod\limits_{p=2}^{k-2}(J_kl_{p,k-1}-l_{1k}-J_k)\over\prod\limits_{p=2}^{k-1}
(J_kl_{pk}-l_{1k})(J_kl_{pk}-l_{1k}-J_k)}\right\}^{{1\over2}}\times\\
\times\left\{-{\matrix
(l_{1,k+1}-l_{1k}j_{n-k})(l_{k+1,k+1}-l_{1k}j_{n-k})
\times\\
\times(l_{1,k-1}j_{n-k+1}-l_{1k}-J_k)\endmatrix
\over(l_{kk}-l_{1k})(l_{kk}-l_{1k}-J_k)(l_{k-1,k-1}j_{n-k+1}-l_{1k}-J_k)^{-1}}
\right\}^{{1\over2}},\endgather
$$
$$
\gather
\widetilde{b}^s_k(m)=\left\{{\prod\limits_{p=2}^k(l_{p,k+1}-l_{sk})
\prod\limits_{p=2}^{k-2}(l_{p,k-1}-l_{sk}-1)\over\prod\limits^{k-1}\Sb
p=2\\ p\neq s\endSb
(l_{pk}-l_{sk})(l_{pk}-l_{sk}-1)}\right\}^{{1\over2}}\times\\
\times\left\{-{(l_{1,k+1}-J_{k+1}l_{sk})\over
(l_{1k}-J_kl_{sk})}\right.\times\\
\times\left.{\matrix
(l_{k+1,k+1}-J_{k+1}l_{sk})(l_{1,k-1}-J_{k-1}l_{sk}-J_{k-1})
\times\\
\times(l_{k-1,k-1}-J_{k-1}l_{sk}-J_{k-1})\endmatrix
\over(l_{1k}-J_kl_{sk}-J_k)(l_{kk}-J_kl_{sk})(l_{kk}-J_kl_{sk}-J_k)}
\right\}^{{1\over2}},\quad 1<s<k,\\
\endgather
$$
The expression for $\widetilde{a}^k_k(m)$ can be derived from that for
$\widetilde{a}_k^1(m)$ by changing $l_{1k}$ for $l_{kk}$ and $l_{kk}$
for $l_{1k}$. The same substitution turns $\widetilde{b}_k^1$ into
$\widetilde{b}_k^k(m)$. Components $m$ are related with components $l$
by equations
$$
l_{1k}=m_{1k}-J_k,\quad l_{kk}=m_{kk}-kJ_k,\quad l_{sk}=m_{sk}-s,
\quad 1<s<k.
\tag 55
$$

As it can be shown by direct checking, operators (53) satisfy
commutation relations (28) of algebra $u\,(n;\boldkey
j)$. Therefore, they give a representation of algebra. Considering
the action of rising operators $A_{n-k,n-k-1}$ on vector of the
``major weight'' $\varphi_{\scriptscriptstyle\text{Mw}}$, described by
scheme $|\,m\rangle$ for maximal values of components, and the action
of lowering operators $A_{n-k-1,n-k}$ on vector of the ``minor
weight'' $\varphi_{\scriptscriptstyle\text{mw}}$, described by scheme
$|\,m\rangle$ for minimal values of components, as in $\S$2.1, we
find that for dual or imaginary values of all or some parameters
$\boldkey j$ the space of representation is infinite-dimensional and
does not contain subspaces invariant in respect to operators (53),
because taking any basis vector and acting on it by operators $A_{kr}$
required number of times, we obtain all basis vectors in the space of
representation. Therefore, representation (53) is irreducible.

Though the initial representation of algebra $u\,(n)$ is Hermitean,
irreducible representation (53) of algebra $u\,(n;\boldkey j)$, in
general, is not Hermitean. Therefore, if we want representation (53)
to be Hermitien, it is necessary to require the fulfilment of
relations $A_{pp}^+=A_{pp}$ $(p=0,1,\dots,n-1)$,
$A_{kp}=A^+_{pk}$, which for matrix elements of operators can be
written as follows:
$$
\gathered
\langle m\,|\,A_{pp}\,|\,m\rangle=\overline{\langle
m\,|\,A_{pp}\,|\,m\rangle},\\
\langle n\,|\,A_{kp}\,|\,m\rangle=\overline{\langle
m\,|\,A_{pk}\,|\,n\rangle},
\endgathered
\tag 56
$$
where bar means complex conjugation.
\bigskip

{\bf 3.2 Spectrum of Casimir operators}
\medskip

Components $m^*_{kn}$ of the upper row of scheme (50) (components of
the highest weight) completely determine the irreducible
representation of algebra $u\,(n)$.  A.M.Perelomov and V.S.Popov
[3], A.N.Leznov, I.A.Malkin, V.I.Man'ko [5] found eigenvalues of
Casimir operators, expressing them in terms of components of the major
weight.  For unitary algebra $u\,(n)$ spectrum of Casimir operators can be
written as follows:
$$
C^*_q(\boldkey m^*)=\Tr a^{*q}E,
\tag 57
$$
where $E$ is matrix of dimension $n$, all elements of which are equal
to unit, and matrix $a^*$ is as follows
$$
a^*_{ps}=(m^*_{pn}+n-p)\delta_{ps}-\theta_{sp},\quad s,p=1,2,\dots,n.
\tag 58
$$
Here $\theta_{sp}=1$ for $s<p$ and $\theta_{sp}=0$ for
$s\geq p$.

Under transition from algebra $u\,(n)$ to algebra
$u\,(n;\boldkey j)$, $\boldkey j=(j_1,j_2,\mathbreak\dots,j_{n-1})$
the components of
major weight are transformed according to (51), i.e.
$m_{1n}=Jm^*_{1n}$, $m_{nn}=Jm^*_{nn}$, $m_{sn}=m^*_{sn}$
$(s=2,3,\dots,n-1)$, $J=\prod\limits_{l=1}^{n-1}j_l$.
Let us define matrix $a(\boldkey j)$ as
$$
a\,(\boldkey j)=Ja^*(\to),
\tag 59
$$
where $a^*(\to)$ is matrix (58), in which the
components $m^*_{pn}$ are substituted by their expressions in terms of
$m_{pn}$, i.e. $a_{11}(\to)=n-1+m_{1n}J^{-1}$,
$a_{nn}(\to)=m_{nn}J^{-1}$, and the rest matrix elements are given by
(58).  Then matrix $a(\boldkey j)$ is as follows:
$$
\gathered
a_{11}(\boldkey j)=m_{1m}+J(n-1),\quad a_{nn}(\boldkey j)=m_{nn}\\
a_{ps}(\boldkey j)=J[(m_{pn}+n-p)\delta_{ps}-\theta_{sp}],\\
p,s=2,3,\dots,n-2.
\endgathered
\tag 60
$$

Casimir operators are transformed according to [2]:
$C_{2k}(\boldkey j)=\mathbreak=J^{2k}C^*_{2k}(\to)$,
$C_{2k+1}(\boldkey j)=J^{2k}C^*_{2k+1}(\to)$.  Their spectraare
transformed in just the same way. Therefore, spectrum of Casimir
operators for algebra $u\,(n;\boldkey j)$ is as follows:
$$ \gathered
C_{2k}(\boldkey m)=J^{2k}\Tr\,
(a^*(\to))^{2k}E=\\=\Tr\,(Ja^*(\to))^{2k}E=\Tr a(\boldkey j)E,\\
C_{2k+1}(\boldkey m)=J^{2k}\Tr\,(a^*(\to))^{2k+1}E=\\
=\Tr\,(a^*(\to)Ja^*(\to))^{2k}E=\Tr a^*(\to)a^{2k}(\boldkey j)E,
\endgathered
\tag 61
$$
where $2k$ and $2k+1$ takes all integer values from 1 to $n$. In
particular,
$$
\gathered
C_1(\boldkey m)=(m_{1n}+m_{nn})J^{-1}+\sum_{s=2}^{n-1}m_{sn},\\
C_2(\boldkey m)=m_{1n}^2+m_{nn}^2+J(n-1)(m_{1n}-m_{nn})+\\
+J^2\sum_{s=2}^{n-1}m_{sn}(m_{sn}+n+1-2s)
\endgathered
\tag 62
$$
are eigenvalues of the first two Casimir operators of
algebra $u\,(n;\boldkey j)$ on the irreducible representation.
\bigskip

{\bf 3.3. Possible variants of contractions}

{\bf of irreducible representations}
\medskip

For brevity, in this section we shall talk on contractions of
irreducible representations, however, keeping in mind that the
corresponding considerations are valid for imaginary values of
parameters $\boldkey j$ as well. Transformation (51) of components
in Gel'fand-Zetlin scheme have been chosen in such a way that
eigenvalues of Casimir operators of even order would differ from zero
under contractions. However, variant (51) (we call it basic) is not
unique. As it can be easily seen from (61), (62), the same goal
can be achieved, transforming any two components of the upper row
according to the rule $m=Jm^*$ and leaving unchanged the other
components of this row.

What happens in this case with initial irreducible representation, say under
contraction
$j_1=\iota_1$, $j_2=\dots=j_{n-1}=1$?
The transformation rule for generators remains unchanged:
$A=(\prod\limits_{k}j_k)A^*(\to)$,
only expressions $A^*(\to)$ for singularly transformed operators of
representation are modified as well as inequalities for components of
Gel'fand-Zetlin scheme in comparison with the same contraction
$j_1=\iota_1$ in basic variant. The eigenvalues of Casimir operators
depend not on components $m_{1n}$, $m_{nn}$, as in basic variant, but
on other two components of the upper row.

Thus, each of
$\pmatrix n\\2\endpmatrix=n(n-1)/2$
variants of transition from irreducible representation of
algebra $u\,(n)$ gives under contractions its own irreducible
representation of algebra $u\,(n;\boldkey j)$, which spectrum of
Casimir operators is determined by its own two components of the upper
row of Gel'fand-Zetlin scheme. In this case all variants of transition
are of general type, i.e. lead to non-zero spectrum of all Casimir
operators, even when all parameters $\boldkey j$ take dual values.

The considerations brought above are valid for each algebra
$u\,(k;\boldkey j')$, $k=2,3,\dots,n-1$, in the chain of subalgebras,
described in $\S$3.1, i.e. for each subalgebra there are
$\pmatrix k\\2\endpmatrix=k(k-1)/2$
variants of transition from irreducible representation of subalgebra
$u\,(k)$ to general irreducible representations of subalgebra
$u\,(k;\boldkey j')$. The latter determine Gel'fand-Zetlin basis.
Therefore each of $\pmatrix n\\2\endpmatrix$ variants of transition
from irreducible representation of algebra $u\,(n)$ to irreducible
representations of algebra $u\,(n;\boldkey j)$ can be written in
$N_{n-1}=\sum\limits_{k=2}^{n-1}\pmatrix k\\2\endpmatrix$
different basises, corresponding to different variants of
transformation of Gel'fand-Zetlin scheme components in the rows with
numbers $k=2,3,\dots,n-1$. In the first two sections of this paragraph
we have described basic variant, in which the first and the last
components of the rows with numbers $k=2,3,\dots,n$ undergo
transformation. It is clear that, if necessary, similar relations can
be written for each of
$N_{n}=\sum\limits_{k=2}^{n}\pmatrix k\\2\endpmatrix$
variants.

\vfill\eject
\centerline {\bf References} \medskip
\item{1.} Gel'fand I.M., Zetlin M.L. Finite-dimensional
representations of gro\-up of unimodular matrices / Dokl. of Acad.
Sci.  USSR. Math. series. -- 1950. -- \underbar{71}, No.5.--
P.825--828.
\item{2.} Gromov N.A., Moskaliuk S.S. Special unitary
group in Cayley-Klein spaces. -- Vienna, 1995. - 21p. (Prepr./The
Ervin Schr\"odinger International Institute for Mathematical Physics,
ESI--220).
\item{3.} Perelomov A.M., Popov V.S. Casimir operators for groups
$U(n)$ and $SU(n)$ // Nucl. Phys. -- 1966. -- \underbar{3}, No.5. --
P.924--930.
\item{4.} Perelomov A.M., Popov V.S. Casimir operators for
orthogonal and symplectic groups  // Ibid. --
1966. -- \underbar{3}, No.6. -- P.1127--1134.
\item{5.} Leznov A.N., Malkin I.A., Man'ko V.I. Canonical
transformations and theory of representations of Lie groups // Proc. of
Phys. Inst. Acad. USSR. -- 1977. -- \underbar{96}. -- P.27--71.
\item{6.} Gromov N.A., Moskaliuk S.S. Special orthogonal groups in
Cayley-Klein spaces. -- Vienna, 1995. -- 33p. (preprint // The Ervin
Schr\"o\-din\-ger International Institute for Mathematical Physics,
ESI--222).
\item{7.} Chakrabarti A. Class of representations of the $iu(n)$
and $io(n)$ algebras and respective deformations to $U(n,1)$, $o(n,1)$
 // J. Math. Phys. -- 1968. -- \underbar{9}, No.12. -- P.2087--2100.
\item{8.} Gel'fand I.M., Graev M.I. Finite-dimensional irreducible
representations of unitary and general linear groups and connected
with them special functions // Izv. of Acad. Sci. USSR. -- Math. ser.
-- 1965. -- \underbar{29}, No.5. -- P.1329--1356.
\item{9.} Linblad G., Nagel B. Continuous bases for unitary
irreducible representations of $SU(1,1)$ // Ann. Inst. H. Poincar\'e.
-- 1970. -- \underbar{13}, No.1. -- P.27--56.
\item{10.} Gel'fand I.M., Graev M.I. Irreducible representations of
Lie algebra of group $U(p,q)$ // Physics of high energy and theory of
elementary particles. -- Kiev: Naukova Dumka, 1967. -- P.216-226.
\item{11.} Barut A., Ronczka P. Theory of group representations and
its applications. -- Moscow: Mir, 1980. -- Vol.1. -- 456p., Vol.2. --
396p.
\item{12.} Celeghini E., Tarlini M. Contractions of group
  representations. I // Nuovo Cim. B. -- 1981. -- \underbar{61}, No.2.
  -- P.265--277.
\item{13.} Celeghini E., Tarlini M. Contractions of group
  representations. II // Ibid. -- 1981. -- \underbar{65},
  No.1. -- P.172--180.
\item{14.} Celeghini E., Tarlini M. Contractions of group
  representations. III // Ibid. -- 1982. --
  \underbar{68},
  No.1. -- P.133--141.
\item{15.} Mukunda N. Unitary representations of the group $O(2,1)$
in an \linebreak $O(1,1)$ basis // J. Math. Phys. -- 1967. --
  \underbar{8},
  No.11. -- P.2210--2220.
\item{16.} Kuriyan J.G., Mukunda N., Sudarshan E.G. Master analytic
representations: reduction of $O(2,1)$ in an $O(1,1)$ basis // J. Math
Phys. -- 1968.
  \underbar{9},
  No.12. -- P.2100--2108.
\item{17.} Inonu E., Wigner E.P. On the contraction of group and
their representation // Proc. Nat. Acad. Sci. USA. -- 1953. --
  \underbar{39},
  No.3. -- P.510-524.
\end